\documentclass[linenumbers]{aastex631}

\def\reference{In: Pei, J., Tseng, V.S., Cao, L., Motoda, H., Xu, G. (eds) Advances in Knowledge
Discovery and Data Mining. Lecture Notes in Computer Science(), vol 7819. Springer, Berlin,
Heidelberg.}

\usepackage{color}

\shorttitle{The SMC star cluster HW~42 }
\shortauthors{Andr\'es E. Piatti}

\begin{document}

\title{Small Magellanic Cloud field stars meddling in star cluster age estimates }

\author[0000-0002-8679-0589]{Andr\'es E. Piatti}
\affiliation{Instituto Interdisciplinario de Ciencias B\'asicas (ICB), CONICET-UNCUYO, Padre J. Contreras 1300, M5502JMA, Mendoza, Argentina}
\affiliation{Consejo Nacional de Investigaciones Cient\'{\i}ficas y T\'ecnicas (CONICET), Godoy Cruz 2290, C1425FQB,  Buenos Aires, Argentina}
\correspondingauthor{Andr\'es E. Piatti}
\email{e-mail: andres.piatti@unc.edu.ar}

\begin{abstract}
I revisited the age of the Small Magellanic Cloud cluster HW~42, whose previous 
estimates differ in more than 6 Gyr, thus challenging the most updated knowledge of the SMC 
star formation history. I performed an analysis of number stellar density profiles at
different brightness levels; carried out a field star decontamination of the
cluster color-magnitude diagram; and estimated the cluster fundamental parameters from
the minimization of likelihood functions and their uncertainties from standard bootstrap 
methods. I conclude that HW~42 is a 6.2$^{\rm +1.6}_{\rm -1.3}$ Gyr old ([Fe/H] = 
-0.89$^{\rm +0.10}_{\rm -0.11}$ dex) SMC cluster
projected on to a SMC composite star field population which shows variations in magnitude, 
color, and stellar density of Main Sequence stars. The present outcome solves the
conundrum of the previous age discrepancies and moves HW~42 to a region in the
SMC age-metallicity relationship populated by star clusters. 
\end{abstract}


\section{Introduction}
There is
a general consensus that the oldest SMC star cluster population
has ages $\sim$ 6-10 Gyr
\citep[][and references thereis]{narlochetal2021}.
HW~42 is a SMC star cluster added to the group of 
intermediate-age star clusters by \citet{p11c}, and latter analyzed by \citet{perrenetal2017}, 
who estimated an age of 5.0 Gyr from relatively shallow photometric 
data. Recently, \citet{bicaetal2022} derived a much younger age (2.6 Gyr), moving
it toward an extreme position in the SMC star cluster age-metallicity 
relationship. The curious younger age and more metal-rich content ([Fe/H]=-0.57) of 
HW~42 caught our attention.



\section{Data analysis and discussion}

\begin{figure}
\includegraphics[width=\columnwidth]{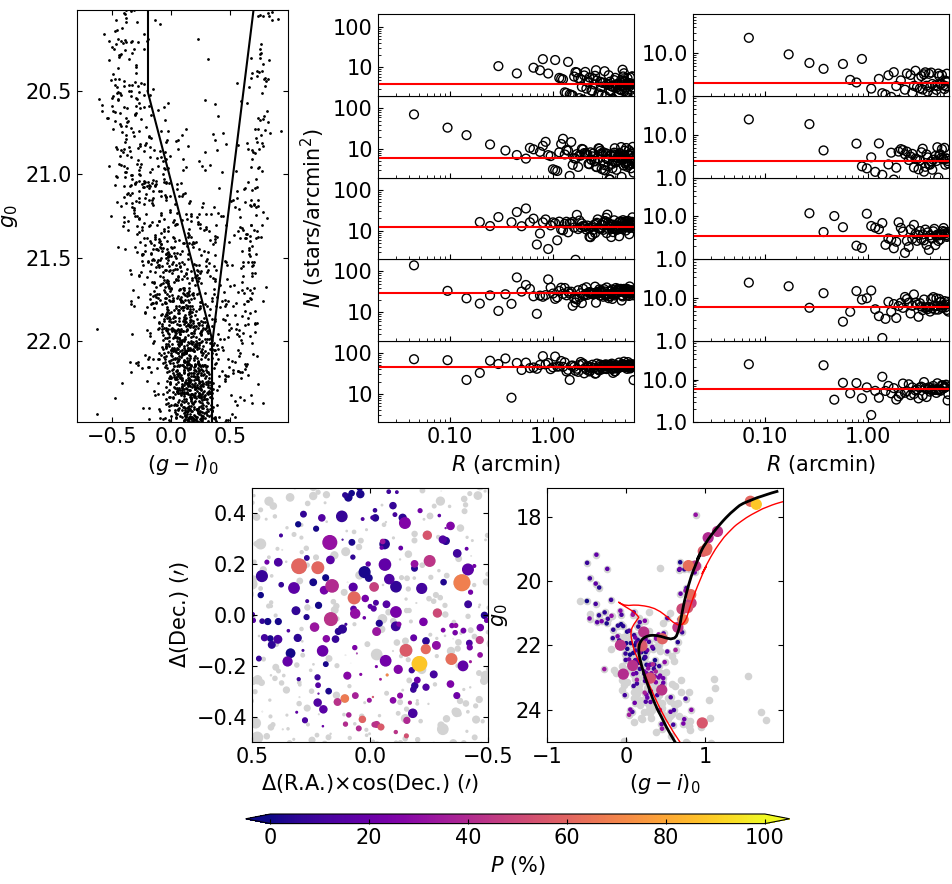}
\caption{{\it Top:} CMD of stars located around a radius of 2' from the
cluster center (left panel) and number density profiles 
for different magnitude bins and for MS (middle panel) and RGB
(right panel) stars. Red lines represent the mean background level.
{\it Bottom:} Chart of all stars measured in the field of HW~42 of its CMD.
The size of the symbols in the chart is proportional to the $g_0$ brightness of the star.
Symbols in both
panels are color-coded according to the assigned membership probability $P$.
Theoretical isochrones in \citet{bicaetal2022} and in this work are superimposed in the CMD
with red and black lines, respectively.}
\label{fig}
\end{figure}

In order to provide an independent estimate of the HW~42's age,
I chose the publicly available Survey of the Magellanic Stellar History
(SMASH) DR2 data sets \citep{nideveretal2021}.
Since SMASH provides individual $E(B-V)$ values, I employed a dereddened $g_0$ versus $(g-i)_0$ CMD.
After extensively examining the cluster CMD I found that observed 
Main Sequence (MS) and Red Giant Branch (RGB) stars have distinguishable spatial distributions.
I arrive at this conclusion by building stellar number density profiles for
stars at five different magnitude interval of 0.5 mag wide from 20.0 down to 22.5 mag
and located to the left or to the right of the lines drawn in Fig.~\ref{fig}.

A close inspection of Fig.~\ref{fig} reveals that there is some excess of MS stars
inside a radius of $\sim$0.5', only for stars with 20.5 $<$ $g_0$ (mag) $<$ 21.0.
For the remaining $g_0$ magnitude bins, number density dispersion around the mean background
level or lack of stars inside a radius of $\sim$0.5' prevail.
If the excess of stars at 20.5 $<$ $g_0$ (mag) $<$ 21.0 corresponds to the cluster stellar
population, and I assume that they are at the cluster MS turnoff, then HW~42 should be as young
as 2.0 Gyr old. However, I would also expect to find fainter excesses of cluster MS stars, which 
is not the case. Additionally, Fig.~\ref{fig} shows excesses of RGB stars mostly along the 
whole magnitude range and inside a radius of $\sim$0.50', which suggest the existence 
of an older stellar aggregate. Furthermore, it is not possible to reconcile a cluster
MS turnoff as bright as $g_0$ $\sim$ 20.5-21.0 mag with a lower end of a cluster RGB
located at $g_0$ $\sim$ 22.5 mag \citep{betal12}. From Fig.~\ref{fig} I conclude 
that I are dealing with the SMC MS field star contamination that misleads the reliable 
identification of the cluster MS turnoff. Indeed, I used to recognize star members of HW~42 the
widely recommended HDBSCAN 
\citep[Hierarchical Density-Based Spatial Clustering of Applications with Noise,][]{campelloetal2013}
Gaussian mixture model technique \citep{hr2021}, with SMASH coordinates as cluster searching variables,
and the resulting most probable group includes MS field stars.


In order to confidently decontaminate the cluster CMD from field stars, I applied
the procedure devised by  \citet{pb12}, which has been shown to produce cleaned cluster CMDs
\citep[e.g.,][and references therein]{pl2022,piatti2022}. 
The cleaning procedure was applied a thousand times, each
execution with a different randomly selected reference field star region. From all the
produced cleaned cluster CMDs, I defined the membership probability $P$ ($\%$) =
$N$/10, where $N$  represents the number of times a star was not subtracted during the 
thousand different CMD cleaning executions. 
Fig.~\ref{fig} shows the spatial 
distribution and the CMD of stars with $P$ $>$ 0$\%$. 
As can be seen, a relatively long
cluster RGB is clearly visible, alongside some few sub-giant and MS stars with $P$ $>$ 50$\%$.


I used theoretical isochrones  computed by 
\citet[][PARSEC\footnote{http://stev.oapd.inaf.it/cgi-bin/cmd}]{betal12} for the
SMASH photometric system to produce synthetic CMDs. PARSEC v1.2S isochrones spanned log(age /yr) 
from 9.0 up to 9.9 in steps of 0.025, and metallicities (log($Z/Z_\odot$)) from 0.0005 
dex up to 0.005 dex, in steps of 0.001 dex. The Automated Stellar Cluster Analysis code 
\citep[\texttt{ASteCA,}][]{pvp15} was employed to simultaneously
derive the metallicity, the age, and the distance of HW~42,
which resulted to be those corresponding to the isochrone associated to the synthetic CMD that best 
matched the cleaned dereddened cluster CMD ($P$ $>$ 50$\%$). 
The uncertainties in the derived parameters were estimated from the standard bootstrap method 
\citep{efron1982}.  The resulting astrophysical properties of HW~42 turned out to be:
distance = 54.45$^{\rm +2.83}_{\rm -2.69}$ kpc (true distance modulus = 
18.68$\pm$0.11 mag); age = 6.2$^{\rm +1.6}_{\rm -1.3}$ Gyr; and [Fe/H] = -0.89$^{\rm +0.10}_{\rm -0.11}$ 
dex. 

The cleaned cluster CMD shows a numerous population of MS stars 
- there are also some sub-giant and giant stars - with $P$ $<$ 10$\%$. These stars
belong to the SMC field star population, and remain in the cleaned cluster CMD
because of their not homogeneous distribution in either magnitude, color, stellar density
or all of these three quantities combined across the cluster field of view. 
As can be seen, HW~42 (stars with distance to the cluster center $<$ 0.5') is 
projected on to a composite field population with MS stars spanning an age range of 
$\sim$1-6 Gyr. Such an SMC MS field star contamination is also seen in the cluster
CMD built by \citet{bicaetal2022}. 
Fig.~\ref{fig}
shows that the cluster MS turnoff arising from this work is fainter than that adopted by
\citet{bicaetal2022}

\begin{acknowledgments}

This research uses services or data provided by the Astro Data Lab at NSF's National 
Optical-Infrared Astronomy Research Laboratory. NSF's OIR Lab is operated by the 
Association of Universities for Research in Astronomy (AURA), Inc. under a cooperative
agreement with the National Science Foundation.
\end{acknowledgments}


\end{document}